\documentclass[12pt]{article}
\usepackage{array,epic,graphpap,graphics,float,tabularx}
\usepackage{amsmath}
\usepackage{amstext}
\usepackage{amsbsy}
\usepackage{amsthm}
\usepackage{amsfonts}
\usepackage{amssymb}
\usepackage{amscd}
\usepackage{eucal}
\usepackage{caption}
\usepackage{url}
\usepackage{color}

\usepackage[round,comma,authoryear,sort&compress]{natbib} 

\theoremstyle{plain}

\theoremstyle{definition}

\numberwithin{equation}{section}
\def\eqdef{\triangleq}
\def\ito{It{\^o}}
\def\d{\mathrm{d}}
\def\a{\alpha}

\def\p{\pi}

\def\s{\sigma}

\def\g{\gamma}

\def\half{\frac{1}{2}}

\def\intT{\int_0^T}
\def\limT#1{\lim_{T\to\infty}\frac{#1}{T}}

\def\as{\quad\text{\rm a.s.}}
\def\sumi{\sum_{i=1}^n}

\def\bg{{\boldsymbol g}}
\def\1{{\bf 1}}
\def\E{{\mathbb E}}

\begin{document}

\title{\bf Diversification, Volatility, and Surprising Alpha}
\author{Adrian Banner\footnote{Intech, 1 Palmer Square, Princeton, NJ 08542.}  \and
Robert Fernholz$^*$ \and Vassilios Papathanakos$^*$ \and Johannes Ruf\footnote{Department of Mathematics,   London School of Economics and Political Science, London WC2A 2AE.}  \and David Schofield\footnote{Intech, 201 Bishopsgate, London EC2M 3AE.}  }
\date{\today}
\maketitle
\begin{abstract}

It has been widely observed that capitalization-weighted indexes can be beaten by surprisingly simple, systematic investment strategies. Indeed, in the U.S.~stock market, equal-weighted portfolios, random-weighted portfolios, and other na\"ive, non-optimized portfolios tend to outperform a capitalization-weighted index over the long term. This outperformance is generally attributed to beneficial factor exposures. Here, we provide a deeper, more general explanation of this phenomenon by decomposing portfolio log-returns into an average growth and an excess growth component. Using a rank-based empirical study we argue that the excess growth component plays the major role in explaining the outperformance of na\"ive portfolios. In particular, individual stock growth rates are not as critical as  is traditionally assumed.
\end{abstract}

\section{Introduction}

In the Summer of 2013 a paper published in the Journal of Portfolio Management entitled `The Surprising Alpha From Malkiel's Monkey and Upside-Down Strategies' by Rob Arnott, Jason Hsu, Vitali Kalesnik and Phil Tindall observed that in the US and Global stock markets equal-weighted portfolios, random-weighted portfolios and other na\"ive, non-optimised portfolios tend to outperform a capitalization-weighted index in the long run. This was a prominent paper which attracted a good deal of attention at the time in both trade and popular press, and won the 2013 Bernstein Fabozzi/Jacobs Levy Award for Outstanding Paper in the Journal of Portfolio Management. 

The apparent fact that the cap-weighted index could so easily be beaten was characterised by the authors as `surprising', `perplexing', `paradoxical' and a `puzzle', and the paper offered by way of explanation two main lessons to be learnt: 1.) the investment thesis underlying the various different portfolios examined was not responsible for the observed outperformance; and 2.) all the portfolios displayed significant size and value factor biases which were credited with explaining most of the outperformance. In the small number of cases for which the extended four-factor risk model was not sufficient to explain all of the observed outperformance, the call was raised to discover other factors to explain it: `Let the quest for the missing risk factor(s) begin!'

Risk factors, especially the `big 4' (market, size, value and momentum) have been adopted by many investment practitioners and finance academics as the basic principal components used to explain portfolio performance. Once it has been established that a portfolio's relative performance is explained by, say, the presence of size and value factors, then no further explanation is thought to be necessary, or even possible as the factors cannot be further broken down.

So prevalent has this mind-set become that any portfolio of which the performance cannot be explained by these 4 factors is thought to indicate the presence of some yet-to-be discovered factor, or the similarly elusive dark-matter of manager skill. Factors are the `atoms' of attribution, the ultimate particles of portfolio performance. 

But of course it is well-known that scientists of the late nineteenth and early twentieth centuries demonstrated that the atom was not the ultimate, indivisible particle of matter -- it could be further decomposed providing one had the right detection equipment.

This paper does not set out to discover the `missing' factor sought by \cite{Arnott:2013}  but we will instead propose an alternative, scientific decomposition for the results  observed in that paper.
Furthermore this decomposition is universally applicable to all portfolios. We repeat a representative sample of the experiments conducted in  \cite{Arnott:2013}  and we explain the results using simple methods first introduced by  \cite{Fernholz:Shay:1982}. The detection equipment used in this case is just mathematics, and the particular lens applied is that of Stochastic Portfolio Theory. 

It has long been known in this field that the cap-weighted portfolio is relatively easy to outperform.  Based on these same methods, precisely structured `na\"ive' portfolios that systematically outperform capitalization-weighted benchmarks were introduced in the 1990s with \cite{Fernholz:Garvy:Hannon}. The theory behind all these methods was reviewed in Fernholz (2002) and more current presentations can be found in \cite{FK_survey}, and \cite{Karatzas:Ruf:2016}.

\section{A review of some basic concepts} \label{S:2_new}
Before describing our experiment and its results, it will be important to review and define various basic concepts that will be crucial to understanding and interpreting these results. 

As we will be examining and attempting to account for the returns of various different portfolios it is important first of all to know exactly what we mean when we talk about return, and what kind of return we are talking about. This may seem trivial but it will ultimately bring to light an important aspect of the long term returns of portfolios.

The classical definition of the return on an investment is simply the difference between the final value and the initial value, divided by the initial value:
\begin{equation*} 
\text{return} \eqdef \frac{\text{final value }-\text{ initial value}}{\text{initial value}}.
\end{equation*}

This calculation is fine for a single-period return but suppose we wish to calculate the average annual return of an investment over several years. Suppose that over $N$ years, a stock has annual returns of $r_1,r_2,\ldots,r_N$. There are several common methods for calculating this and they all have different characteristics:

\begin{enumerate}

\item {\em Arithmetic average return:} This is simply calculated as the sum of all the annual returns, divided by the number of years:
\[
\frac{1}{N}\Big((1+r_1)+\cdots+(1+r_N)\Big)-1.
\]
This form of return is widely used in Modern Portfolio Theory and is compatible with the linear models used to calculate the Sharpe ratio and beta. It is, however, upward-biased as an estimator of expected long-term growth and can lead to absurd estimates in some cases. For example consider the case when a +100\% return one year is followed by a -50\% return the following year. Here the average arithmetic return over the two-year period is 25\%, whereas in reality such an investment would have zero growth over the two years.

\vspace{8pt}
\item {\em Geometric average return:} 
For a period of $N$ years, this is calculated as the $N$-th root of the product of the annual returns:
\[
	 \sqrt[N]{(1+r_1)\times\cdots\times(1+r_N)}-1. 
\]
This form of return may be the most common method in practice. It helps to avoid the absurd results apparent in the example of arithmetic average returns given above, and this gives the method a somewhat scientific gloss. Unfortunately the geometric return is awkward to work with, compatible with neither the Sharpe ratio nor beta, and it too is upward-biased as an estimator of expected long-term growth.

\vspace{8pt}
\item {\em Logarithmic average return:} This is calculated simply as the sum of the logarithms of the annual returns, divided by the number of years:
\[
	\displaystyle \frac{1}{N}\Big(\log(1+r_1)+\cdots+\log(1+r_N)\Big). 
\]
Logarithmic return is used in Stochastic Portfolio Theory and is the only one of these three alternative measures of average return that is \emph{unbiased} as an estimator of expected long term growth. 
\end{enumerate}

It can be seen from these definitions that\footnote{Mathematically, these inequalities can be proven via an application of Jensen's inequality.}
\begin{equation*}
\text{ arithmetic return } \ge \text{ geometric return } \ge \text{ logarithmic return}.
\end{equation*}

For the remainder of this paper we shall concentrate on arithmetic and logarithmic returns, and shall refer to arithmetic return simply as `return' and logarithmic return as `log-return'.

\section{The relationship between return and log-return} \label{S:3_new}
For any single stock there is a now well-known relationship between the return of the stock and its log-return as follows:\footnote{For a more detailed discussion and derivation of this relationship see Appendix~\ref{S:2.2}, `The dynamics of arithmetic return and log-return'. }
\[
	\text{log-return of stock} \approx \text{return of stock} - \frac{\text{variance of return}}{2}.
\]
In other words the log-return of the stock is approximately equal to its arithmetic return less half its variance. This latter term is often referred to as the `volatility drag', the negative impact on a stock's long term compound growth arising from its volatility. This was noticed in \cite{Fernholz:Shay:1982}.

\section{Portfolio return and log-return} \label{S:4_new}
Until now we have been considering the relationship between different measures of return for single stocks. We shall now apply this to portfolios. 

As one might expect, the return of a portfolio over a single period is simply the weighted average return of all the stocks making up the portfolio. This was first formalised in \cite{Markowitz:1952}, however the same does not apply for a portfolio's log-return.\footnote{For a more mathematical discussion of the results in this section, see Appendix~\ref{A:B}, `Portfolio return and log-return -- the mathematics'.}

When applying the relationship between return and log-return for a single stock, given in the previous section, to a portfolio comprised of multiple stocks, it emerges that the log-return of a portfolio is not simply the weighted average log return of its constituents -- remarkably it is actually greater than that:
\[
\text{portfolio log-return} = \text{weighted average stock log-return} + \text{excess growth rate}.
\]

The amount by which a portfolio's log-return exceeds that of its stocks is known in the literature of Stochastic Portfolio Theory as a portfolio's excess growth rate, and was first noted in  \cite{Fernholz:Shay:1982}.

The excess growth rate (EGR) itself is simply defined as follows:
\[
\text{EGR} = \frac{\text{weighted average stock variance} - \text{portfolio variance}}{2}.
\]

For practical purposes, given the above definitions, it can be seen that the excess growth rate is an important component of a portfolio's log-return. It measures the positive boost to a portfolio's long-term return that arises from the extent to which the volatility of the portfolio is less than that of its constituent stocks. That is to say, it represents the boost to return that arises from the efficacy of diversification.

Importantly it can even be shown that this quantity cannot be negative for a long-only portfolio (see  \cite{Fe}). It is also clear to see that the excess growth rate will be higher for portfolios of volatile stocks with low correlations. \emph{If all else is equal, a higher excess growth rate will increase the long-term growth of a portfolio.}

\section{Estimation of expected portfolio log-return with a rank-based stock analysis}  \label{S:3}

To recap, a portfolio's log-return can be decomposed into two key components:
\[
\text{portfolio log-return} = \text{weighted average stock log-return} + \text{excess growth rate}.
\]
We can now use this natural decomposition to estimate the expected log-returns of portfolios. For convenience we shall refer to the weighted average stock log-return as the average growth component, and the excess growth rate as the excess growth component.

The excess growth component can be estimated relatively easily, since its value depends only on variances, or relative variances, which are not difficult to determine in practice. 

The average growth component, however, is more difficult to estimate. As most aspiring stock-pickers will testify, the expected returns or log-returns of individual stocks are difficult to estimate with any accuracy, and this has been known in the literature since at least \cite{Sharpe:1964}. Fortunately, in the case of our proposed experiment, we do not need to estimate the individual expected log-returns of the individual stocks.

Since we are considering na\"ive strategies for the top 1000 stocks, where the portfolio weights are assigned essentially at random and not picked by a stock-picker, the stock's rank, in terms of its market capitalization, is more important to us than its name. We can therefore use rank-based methods  to determine the value of the average growth component. We do this by measuring the average rank-based log-return (the average log-return over time associated with whichever stock is occupying a given rank in the largest 1000 stocks), without considering stocks individually by name (see \cite{Fe}).\footnote{For a more mathematical discussion of these procedure, see Appendix~\ref{A:C}, `Estimation of expected portfolio log-return with a rank-based stock analysis -- the mathematics'.}

In order to perform this calculation, we selected the 1000 largest stocks by market capitalization on every trading day for the period from 1964 to 2012, and ranked them in order of size, largest to smallest. We then measured the log-returns of the stock at each rank on each trading day and computed the daily average for each of the 1000 ranks, which was finally annualised by multiplying by 250. 

\begin{figure}[H]
\begin{center}
\scalebox{.70}{ \rotatebox{0}{
\includegraphics{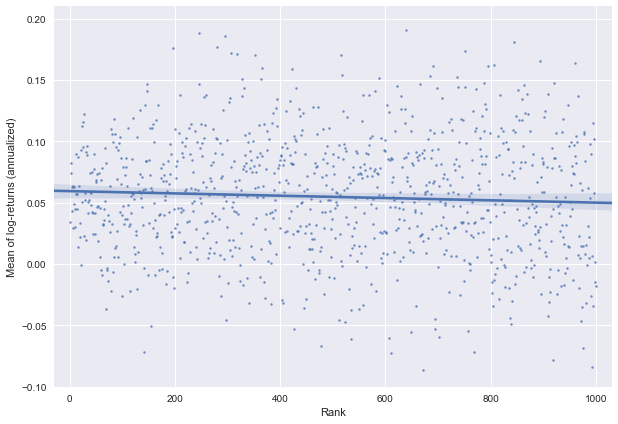}}}
\caption{Estimated rank-based log-returns for top 1000 U.S.~stocks (1964 to 2012). Each trading day, the 1000 largest stocks (measured in terms of market capitalization) are selected and ranked from the largest to the smallest. Then the log-returns of the $k$-largest stocks each day are averaged, yielding 1000 different averages. These averages are then annualised by multiplying them by 250.
This chart contains these averages of log-returns, plotted against the corresponding rank. The slightly decreasing line is a least-squares fit of these points. Its slope is around $-.00001  \pm .00001$ (2 standard errors). See Appendix~\ref{A:data} for a description of the data.}\label{f1}
\end{center}
\end{figure}

The results of this analysis are demonstrated in Figure~\ref{f1}.

The slightly downward-sloping line in Figure~\ref{f1} is a least-squares fit of all the points. Its slope is around $-.00001 \pm .00001$ (2 standard errors).  This would seem to indicate that there is not much difference between the ranked stock growth rates, which implies that the portfolio's expected average growth component should be about the same for all na\"ive portfolios.  If it were the case, for example, that smaller stocks do indeed have higher long term returns then we would expect the line to be upward-sloping, which it manifestly is not.

Given this result it follows that any variation between the log-returns of different portfolios will depend largely on the differences in their respective excess growth components. Furthermore, since the excess growth component depends only on variances, we can conclude that the differences between the log-returns of na\"ive portfolios will depend only on variances and covariances.

\begin{figure}[H]
\begin{center}
\scalebox{.75}{ \rotatebox{0}{
\includegraphics{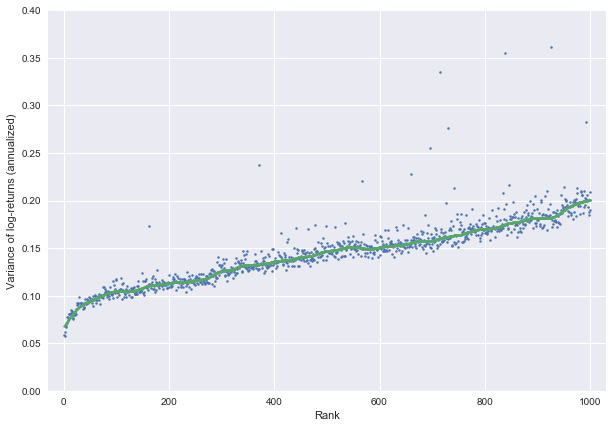}}}
\caption{Annual variance by market capitalization rank for top 1000 U.S.~stocks (1964 to 2012). As in Figure~\ref{f1}, the 1000 largest stocks are ranked from the largest to the smallest each trading day. Then the sample variance of the log-returns of the $k$-largest stocks each day is computed, and annualised. 
This chart contains these sample variances of log-returns, plotted against the corresponding rank. The green line is a smoothed version based on the LOWESS smoother with 5\% of data used.}   \label{f2}
\end{center}
\end{figure}

If we now look at stocks' variances by rank, rather than returns, we see a very different picture. Figure~\ref{f2} confirms that smaller stocks tend to have a larger variance, and since
we have just established that a higher variance leads to a higher excess growth component,
this would lead us to expect that portfolios which are more diversified into smaller stocks
will have a higher return.

However this explanation of the outperformance of smaller stock portfolios is very much at odds with conventional wisdom. In traditional finance, given the assumed positive relationship between risk and return, smaller stocks are expected to have higher returns as compensation for their increased riskiness. These higher expected returns for smaller stocks translate into higher expected returns for portfolios comprised of these smaller stocks.

This may be true for single period arithmetic returns, but as long-term investors we should care about log-returns. We have already demonstrated in Figure~\ref{f1} that the log-returns of the smaller stocks are not, in fact, higher, and so when viewed through the lens of stochastic portfolio theory it becomes
clear that the observed outperformance of small stock portfolios is not due to  higher long-term returns of the small stocks themselves but to a higher portfolio excess growth rate.

\section{The experiments}

\cite{Arnott:2013} test several na\"ive, non-optimized portfolio strategies versus a capitalization-weighted benchmark of the largest 1000 U.S.~stocks over the period 1964-2012. All these strategies have a higher return than the benchmark, and most have a higher Sharpe ratio. It is well-known that capitalization-weighted portfolios are not particularly well-diversified -- there is too much weight concentrated into the largest stocks. All of the na\"ive strategies have more diversification into the smaller stocks than the capitalization-weighted index. 

Importantly, this greater diversification into the smaller stocks is not likely to affect the average growth component much. As we have seen in Figure~\ref{f1}, the growth rates are about the same for the top 1000 stocks. In other words, the na\"ive strategies' outperformance is not due to  higher long-term returns of the smaller stocks.  However the greater diversification is likely to increase the portfolio's excess growth component, since both improved \emph{diversification} and higher stock \emph{volatility} increase excess growth. 

To see what actually happens we run an experiment on the largest 1000 U.S. stocks using overlapping one-year periods starting each month from 1964-2012, quite similar in spirit to the experiment of \cite{Arnott:2013}. 

More precisely, at the beginning of each month we choose the largest 1000 U.S.~stocks and use their one-year returns over the following year to compute the returns of the various strategies described below. Altogether there are 5384 different stocks which were, at the beginning of some month during this 49-year period, among the top 1000 stocks by market capitalization in the U.S.

We do not replicate all the strategies tested by \cite{Arnott:2013} but instead choose 5 representative na\"ive strategies. These 5 buy-and-hold strategies begin each one-year period with the following weights:

\begin{enumerate}
\item {\bf Capitalization-weighted (CW):}  stock weights proportional to their  market capitalization.
\item {\bf Equal-weighted (EW):} weight of each stock = $1/1000$.
\item {\bf Large-overweighted (LO):} stock weights proportional to the square of their market capitalization.
\item {\bf Random-weighted (RW):} weights proportional to $[0,1]$--uniformly distributed random variables.
\item {\bf Inverse-random-weighted (IRW):} weights proportional to the reciprocals of $[0,1]$--uniformly distributed random variables.
\end{enumerate}
All weights are always normalized to sum to 1.

The capitalization-weighted strategy corresponds to holding the market. 

The equal-weighted strategy splits the capital at the beginning of each one-year period equally among whatever the top 1000 names are at that point in time.

The large-overweighted strategy is not tested by \cite{Arnott:2013}. This strategy puts a higher proportion than the index into the larger stocks and a smaller proportion into the smaller stocks.\footnote{
To see this, consider a market with two stocks with relative market capitalizations $\mu$ and $\nu$, where $\nu < \mu$ and $\mu + \nu = 1$. Then $\nu^2 < \mu \nu$; hence the large-overweighted strategy puts the proportion
\begin{align*}
	\frac{\mu^2}{\mu^2 + \nu^2} > \frac{\mu^2}{\mu^2 + \mu  \nu} = \frac{\mu}{\mu +  \nu} = \mu
\end{align*}
of the wealth into the larger stock.}  This portfolio is even less diversified than the capitalization-weighted portfolio, and according to our thesis we would therefore expect it to underperform the market, that is to say, to have negative excess return by virtue of its lower excess growth component. 

The random-weighted and inverse-random-weighted strategies are our version of the  `monkey' and `upside-down' portfolios in \cite{Arnott:2013}. 

To avoid exposure to random draws we simulate 1000 such portfolios and report the median values in Table~\ref{T:1} below.

Each of the five columns in the table corresponds to one of the five strategies described above. We show the total logarithmic return for each strategy, and then decompose the total return into the two components we have discussed extensively in this paper: the average growth component and the excess growth component. In all cases we show both the absolute values as well as the relative values compared to those of the cap-weighted portfolio. Finally we show the arithmetic absolute and relative returns, the standard deviation of the arithmetic returns, and the associated Sharpe ratios.

\vspace{20pt}
\begin{center}
\begin{tabular}{lccccc} 
\bfseries   &   {\bf CW}(\%)  & {\bf EW}(\%)  &  {\bf LO}(\%)  &{\bf RW}(\%) & {\bf IRW}(\%)  \\ 
\hline
{Total log-return} & {9.12}  &  {10.98}  &  {7.46}      &{10.98 }     &{10.46}\\
relative to cap-weighted index & &1.86    &-1.66      &1.86      &1.34\\
\hline
{Average growth component} &{5.57}& {5.64}  &{5.36}   &{5.65}      &{5.67}\\
relative to cap-weighted index &&.07     &-.21     &.08      &.10\\
\hline
\bf{Excess growth component}  & \bf{3.87} & \bf{5.82}   & \bf{2.19}       &\bf{5.82}    &\bf{5.18}\\
relative to cap-weighted index  & &1.95      &-1.68    &1.95      &1.31\\
\hline
Total arithmetic return & 10.97&13.33    &  9.15   &13.33    &13.34\\
relative to cap-weighted index  & &2.36   &   -1.82     &2.36     &2.37\\
\hline
Standard deviation& 17.07&19.14     &16.90       &19.07      &22.35\\
Sharpe ratio&.29&.38     &.18      &.38     &.32\\
\hline
\end{tabular}
\captionof{table}{ Summary of the experiment outcomes. Here, the five columns correspond to the following five strategies, described in the text: CW = capitalization-weighted, EW = equal-weighted, LO = large-overweighted, 
RW = random-weighted,  IRW = inverse-random-weighted.  These strategies are applied, to the largest 1000 U.S.~stocks (at that point of time), month-by-month, from 1964 to 2012, to  using overlapping one-year periods. At the beginning of each period the weights are fixed and not changed over the whole year. The average growth component  and the excess growth component are described at the beginning of Section~\ref{S:3};\protect\footnotemark{ } moreover, the standard deviations of the arithmetic returns are provided.\protect\footnotemark{ }}   \label{T:1}
\end{center}
\vspace{20pt}

\addtocounter{footnote}{-1}

\footnotetext{
The excess growth rate is computed directly, by using the last display of Section~\ref{S:4_new}, or, alternatively \eqref{eq:180512} in the appendix.   This requires computing the covariance matrix of the ranked stocks' returns.  Alternatively, the
 excess growth component could be computed by using the first display in Section~\ref{S:4_new}.
 If the covariance matrix is calculated accurately, which is difficult to do, then these two valuations will be about the same.
 As the reader can check,  these alternative ways of computing the values do not change the table's qualitative conclusion that the excess growth components explain most of the differences in the log-returns.
}

\addtocounter{footnote}{1}

\footnotetext{
For the random-weighted and inverse-random-weighted strategies, Table~\ref{T:1} only reports the median values of 1000 experiments.  Let us provide here some more values. The 10th and 90th percentile of the obtained average growth component values are $5.61$ and $5.68$, and $5.05$ and $6.27$, respectively.  Correspondingly, the 10th and 90th percentile of the obtained excess growth component values are $5.81$ and $5.82$, 
and $5.13$ and $5.23$, respectively. For the Sharpe ratios, the 10th and 90th percentile of the obtained values are $0.38$ and $0.38$, and $0.29$ and $0.35$, respectively.

Any typical experiment involving random weights would provide average and excess growth components that are different from the equally weighted ones. However,  Table~\ref{T:1} reports the median values of many experiments, and these median values turn out to be close to the equally weighted ones.
}

The equal-weighted, random-weighted and inverse-random-weighted portfolios all outperform the capitalization-weighted portfolio. The large-overweighted strategy, on the other hand, underperforms. These results are consistent with our understanding that that the first three strategies have greater diversification than the index, while the latter is less well-diversified. 

Importantly we also note that most of the differences in the strategies' returns can be explained by differences in the excess growth component. Indeed the differences in the average stock growth component are of a much lesser magnitude, which is consistent with the analysis in Figure~\ref{f1} in which we demonstrated that the average individual stock growth rates were essentially the same.

\section{Conclusion }

The outperformance of a range of different strategies of the type presented in  \cite{Arnott:2013} can be explained using concepts from Stochastic Portfolio Theory.

The logarithmic return of a portfolio can be decomposed into two elements. The first term represents the weighted average of the logarithmic returns of the stocks. The second term measures the excess growth, the additional component of portfolio return arising from the benefits of diversification. This term only depends on the variances and covariances of the portfolio's constituents, and is larger for more diversified portfolios. 

Taking a rank-based view of stock returns we argued empirically that the contribution of the weighted average of the logarithmic stock returns is approximately the same for all portfolios. What varies much more from portfolio to portfolio is the excess growth component, which depends only on stocks' variances and covariances. 

Studying the performance of several different trading strategies, some more diversified and some less diversified than the capitalization-weighted portfolio, confirmed these insights. In general, the more diversified portfolios outperform and the single less diversified portfolio underperforms, because the more diversified portfolios have a higher excess growth rate. This arises from the higher variances associated with the smaller stock exposure in these more diversified portfolios, and not because such stocks have inherently higher returns. This higher excess growth rate, in turn, increases the portfolios' logarithmic returns.

All in all, this helps to explain the `surprising' alpha found by \cite{Arnott:2013}  in a variety of strategies, without the need to invoke factors.

\newpage
\appendix
\section*{Appendices}

\section{The dynamics of arithmetic return and log-return} \label{S:2.2}
Let $X(t)$ represent the price of a stock at time $t$. The standard continuous-time model for the behavior of this price is an  \ito\ process that satisfies
\begin{equation*}
\frac{dX(t)}{X(t)}=\a(t) \d t+\s(t) \d W(t),
\end{equation*}
where $\a(t)$ is the {\em rate of return process} of $X$ at time $t$, $\s^2(t)>0$ is the  {\em variance rate process,} and $W$ is a Brownian motion process. For simplicity we assume that $\s^2(t)$ is bounded.

With $X$ as above, \ito's rule (see \cite{KS1}) implies that
\begin{align*}
\d\log X(t) 
&=  \frac{\d X(t)}{X(t)}-\half\s^2(t) \d t\\
&= \Big(\a(t)-\half\s^2(t)\Big)\d t+ \s(t)\,\d W(t)\\
 &=  \g(t)\,\d t + \s(t)\,\d W(t).\phantom{\half}
\end{align*}
The process $\displaystyle{\g(t)=\a(t)-\half\s^2(t)}
$ is called the  {\em growth rate process} (see \cite{Fernholz:Shay:1982}) or the {\em log-return process} of $X$.  This explains the display in Section~\ref{S:3_new}.

Under mild regularity conditions it can be shown that
\begin{equation*}
\limT{1}\Big( \log X(T)- \intT\g(t)\d t\Big)=0,\as
\end{equation*}
This confirms the claim made in Section~\ref{S:2_new} that logarithmic return is an unbiased estimator of long-term growth. Moreover, the fact that $\gamma(t) \leq \alpha(t)$ is consistent with the fact that logarithmic return $\leq$ arithmetic return.

\section{Portfolio return and log-return -- the mathematics}\label{A:B}

In this appendix, we provide the mathematical formulas for the statements in Section~\ref{S:3_new}.

Suppose we have stocks $X_1,\ldots,X_n$ and a portfolio $\p$ with  weights $\p_1(t)+\cdots+\p_n(t)=1$ and value $Z_\p(t)$ at time $t$. Then the portfolio return satisfies
\begin{equation*}
\frac{\d Z_\p(t)}{Z_\p(t)}= \sumi\p_i(t)\,\frac{\d X_i(t)}{X_i(t)}
\end{equation*}
according to \cite{Markowitz:1952}. The analogous equation for the portfolio log-return is
\begin{equation} \label{eq:171103.1}
\d\log Z_\p(t)=\sumi\p_i(t)\d\log X_i(t) + \g^*_\p(t)\d t,
\end{equation}
where $\g_\p^*(t)$ denotes the {\em excess growth rate (EGR) process} of the portfolio. More precisely, if we denote the portfolio variance rate process by $\s^2_\pi(t)$, then we have
\begin{align*}
 \d \log Z_\p(t) &=\frac{\d Z_\p(t)}{Z_\p(t)}-\half\s^2_\p(t)\d t\\
& = \sumi\p_i(t)\,\frac{\d X_i(t)}{X_i(t)}-\half\s^2_\p(t)\d t\\
& = \sumi\p_i(t)\Big(\d\log X_i(t)+\half\s^2_i(t)\,dt\Big)-\half\s^2_\p(t)\d t\\
&= \sumi\p_i(t)\d\log X_i(t)+\half\Big(\sumi\p_i(t)\s^2_i(t) - \s^2_\p(t)\Big)\d t\\
&= \sumi\p_i(t)\d\log X_i(t)+\g^*_\p(t)\d t;
\end{align*}
hence
\begin{equation} \label{eq:180512}
\g^*_\p(t)=\half\Big(\sumi\p_i(t)\s^2_i(t) - \s^2_\p(t)\Big).
\end{equation}
This equality corresponds to the last display in Section~\ref{S:3_new}.

\section{Estimation of expected portfolio log-return with a rank-based stock analysis -- the mathematics}  \label{A:C}
We shall use the notation of Appendix~\ref{A:B}. For the interval $[0,T]$, \eqref{eq:171103.1} yields
\begin{align*}
\text{portfolio log-return } &=\int_{0}^{T} \sumi  \p_i(t)\, \d\log X_i(t)+ \int_{0}^{T} \g^*_\p(t) \d t = \text{A}_\p(T) +\Gamma_\p(T) 
\end{align*}
where 
\begin{equation*}
\text{A}_\p(T) = \int_{0}^{T} \sumi  \p_i(t)\, \d\log X_i(t)
\end{equation*}
is called the {\em average growth} component, representing the weighted average growth rate of the stocks in the portfolio, and
\begin{equation*}
\Gamma_\p(T)  = \int_{0}^{T} \g^*_\p(t) \d t
\end{equation*}
is called the {\em excess growth} component.

To describe the rank-based method mathematically, used to determine the value of the average growth component, 
 let $r_t(i)$ be the rank of $X_i(t)$. That is, if $i$ corresponds to the company with the largest capitalization at time $t$, then $r_t(i) = 1$.  Similarly, if $i$ is the $k$-largest company at time $t$, we have $r_t(i) = k$.  This notation allows us  to define the {\em average rank-based growth rates} $\bg_k$ over $[0,T]$  by
\begin{equation}\label{3}
\bg_k = \frac{1}{T} \int_{0}^{T}\sumi \1_{\{r_t(i)=k\}}\d\log X_i(t).
\end{equation}
In  a stable system the time-averaged value is equal to the expected value, so that
\begin{equation*}
\E\big[\d\log X_i(t) \big| r_t(i)=k\big] \,=\, \bg_k\d t,
\end{equation*}
or 
\begin{equation*}
\E\big[\d\log X_i(t) \big]\,=\, \E\big[\bg_{r_t(i)}]\d t.
\end{equation*}
The definition in \eqref{3} can be used directly to estimate the values of the $\bg_k$, and these estimated values for the period from 1964 to 2012 appear in Figure~\ref{f1} above.

Since the values seem to be roughly the same we shall assume, for the moment, that $\bg_k = \bg$, which then yields
\begin{equation*}
\E\big[\d\log X_i(t) \big]\,=\, \bg \d t.
\end{equation*}

We can now use these values of $\bg_k$ to estimate the expected  average growth component over the period studied. Indeed,
\begin{align}
\E\big[\text{A}_\p(T)\big]
&=  \E\Bigg[ \int_{0}^{T} \sumi  \p_i(t)\d\log X_i(t)\Bigg]\notag\\ 
 &=   \int_{0}^{T} \sumi \E\big[ \p_i(t)\d\log X_i(t)\big]\notag\\
& \simeq \int_{0}^{T}\sumi \E\big[\p_i(t)\big]\E\big[\d\log X_i(t)\big]\label{4}\\
&=  \int_{0}^{T}\sumi\E\big[\p_i(t)\big]\bg \d t,\label{5}
 \end{align}
where the approximate equality in \eqref{4} is justified by the fact that $\p$ is na\"ive, and hence the the weight $\p_i(t)$ and the return on the weight $\d\log X_i(t)$ should be  independent. If this were a portfolio constructed by a skilled stock picker, then we would expect a positive correlation between these two quantities. From Figure~\ref{f1} it appears that there is not much difference among the ranked stock growth rates, so \eqref{5} implies that the portfolio's expected average growth component  should be about the same for all na\"ive portfolios.

\section{Data sources}  \label{A:data}
The stock price data for the figures and the backtesting of the strategies are provided by CRSP. In order to be as close as possible to the experimental setup of \cite{Arnott:2013}, we use data only from 1964 to 2012. However, we have tested all results with data up to 2017, and they are robust.  

There are two returns of -100\% in the dataset.  For the charts of Figures~\ref{f1} and \ref{f2} and for computing $A_\pi$ in Table~\ref{T:1}, these two returns are changed to -95\%.  Otherwise, the corresponding log-returns would be $-\infty$ and the corresponding ranks would not have finite sample averages and variances. The results of this paper are robust with respect to the choice of the number -95\%; other choices would lead to basically the same results.

For a few data points, returns were missing due to incomplete delisting information. We tested the results with different inputs, and the results were robust.

For computing the Sharpe ratio in Table~\ref{T:1} and the necessary excess returns, we used the one-year U.S.~Treasury yields, publicly available on \url{https://www.federalreserve.gov/pubs/feds/2006/200628/200628abs.html}; see also \cite{GSW}.

\bibliographystyle{chicago}
\bibliography{aa_bib}

\end{document}